\documentclass[aps, prb, reprint, lengthcheck, footinbib, showpacs, superscriptaddress, citeautoscript]{revtex4-1}

\pdfoutput=1

\usepackage[T1]{fontenc}
\usepackage[english]{babel}

\usepackage{amsmath}
\usepackage{amssymb}

\usepackage{textcomp}
\usepackage{times}
\usepackage{mathptmx}

\usepackage{braket}
\usepackage{graphicx}

\usepackage[colorlinks, urlcolor=blue, citecolor=blue, linkcolor=blue, pdfstartview=FitH]{hyperref}
\usepackage[all]{hypcap}

\usepackage[dvipsnames]{xcolor}

\begin{document}

\def\affiSOLAB{Spin Optics Laboratory, Saint~Petersburg State University, 198504 St.~Petersburg, Russia}
\def\affiE2a{Experimentelle Physik 2, Technische Universit\"at Dortmund, D-44221 Dortmund, Germany}
\def\affiIOFFE{A.~F. Ioffe Physical-Technical Institute, Russian Academy of Sciences, 194021 St.~Petersburg, Russia}
\def\affiBochum{Angewandte Festk\"orperphysik, Ruhr-Universit\"at Bochum, D-44780 Bochum, Germany}
\def\affiPaderborn{Department Physik, Universit\"at Paderborn, 33098 Paderborn, Germany}

\title{Nuclear magnetic resonances in (In,Ga)As/GaAs quantum dots studied by resonant optical pumping}

\author{M.~S.~Kuznetsova}
\affiliation{\affiSOLAB}
\author{K.~Flisinski}
\affiliation{\affiE2a}
\author{I.~Ya.~Gerlovin}
\author{M.~Yu.~Petrov}
\author{I.~V.~Ignatiev}
\author{S.~Yu.~Verbin}
\affiliation{\affiSOLAB}
\author{D.~R.~Yakovlev}
\affiliation{\affiE2a}
\affiliation{\affiIOFFE}
\author{D.~Reuter}
\affiliation{\affiPaderborn}
\author{A.~D.~Wieck}
\affiliation{\affiBochum}
\author{M.~Bayer}
\affiliation{\affiE2a}

\date{\today}

\begin{abstract}
The photoluminescence polarizations of (In,Ga)As/GaAs quantum dots
annealed at different temperatures are studied as a function of
external magnetic field (Hanle curves). In these dependencies,
remarkable resonant features appear due to all-optical nuclear
magnetic resonances (NMR) for optical excitation with modulated
circular polarization. Application of an additional radio-frequency
field synchronously with the polarization modulation strongly
modifies the NMR features. The resonances can be related to
transitions between different nuclear spin states split by the
strain-induced gradient of the crystal field and by the externally
applied magnetic field. A theoretical model is developed to simulate
quadrupole and Zeeman splittings of the nuclear spins in a strained
quantum dot. Comparison with the experiment allows us to uniquely
identify the observed resonances. The large broadening of the NMR
resonances is attributed to variations of the quadrupole splitting
within the quantum dot volume, which is well described by the model.
\end{abstract}

\pacs{78.67.Hc, 78.47.jd, 76.70.Hb, 73.21.La}

\maketitle

\section*{Introduction}

Nuclear magnetic resonance (NMR) is based on the resonant absorption
of radio frequency (RF) electromagnetic radiation by nuclear spin-split
states by a magnetic field. The absorption of RF~field is limited by
small differences of the populations of the nuclear-spin states in
thermodynamic equilibrium. Therefore NMR requires probing of a
macroscopically large number of nuclei
($\mathord{\sim}10^{18}$)~\cite{Abragam}. The NMR detection
sensitivity can be greatly improved by preparing a nuclear-spin
state with degree of polarization higher than the thermodynamic
equilibrium value at a given magnetic field~\cite{LampelPRL68}.
Additional improvement is achieved by using optical methods to
detect the state of the nuclear-spin system~\cite{Paget-OO,
KKM-SPS}. The efficiency of optically detected NMR was demonstrated
for bulk semiconductors~\cite{EkimovJETPLett72, PagetPRB81,
PagetPRB82}, quantum wells~\cite{KalevichJETPLett90, FlinnJLum90,
KalevichApplMagnRes91, BarrettPRL94, PoggioAPL05}, and even quantum
dots (QDs)~\cite{GammonScience97, DzhioevJETPLett98, MakhoninPRB10,
FlisinskiPRB10, MakhoninNatMat11, ChekhovichNatNano12}, containing
$\mathord{\sim}10^5$ nuclear spins only, thanks to the high values
of dynamic nuclear-spin polarization (DNP), of the order of tens
percent, that can be achieved in QDs by optical
pumping~\cite{GammonPRL01, BraunPRB06, TartakovskiiPRL07,
MaletinskyPRB07, OultonPRL07, CherbuninPRB09, ChekhovichPRL10,
KrebsPRL10, UrbaszekRMP13}.

For self-assembled QDs the observation of NMR is complicated by the
large spread of parameters in a QD ensemble~\cite{FlisinskiPRB10}.
Usually, single QD spectroscopy is used to diminish this
spread~\cite{GammonScience97, MakhoninNatMat11,
ChekhovichNatNano12}. However, a considerable broadening of the
resonances remains even at the single QD
level~\cite{ChekhovichNatNano12}. The main origin of this broadening
in self-assembled QDs is the inhomogeneous quadrupole splitting of
the nuclear-spin states by the strain-induced gradient of crystal
field within a dot~\cite{DzhioevPRL07, MaletinskyNatPhys09}. The
broadening results in overlapping resonances from the same and from
different isotopes, making their identification difficult.

Here, we show that such an identification is nevertheless possible.
We investigate two samples with self-assembled (In,Ga)As/GaAs QDs
grown by the Stranski-Krastanov method and annealed at different
temperatures. The annealing gives rise to a decrease of the
deformation of the crystal lattice, caused by the mismatch of the QD
and barrier lattices and, hence, to a decrease of the inhomogeneous
quadrupole splitting of nuclear states.

We make use of polarization modulation of the optical pumping of the
dots. Such modulation has been the common basis for the all-optical
NMR technique~\cite{KalevichFTT86, KikkawaScience00}. As the
hyperfine interaction of nuclei with oriented electron spins can be
described as an effective magnetic field (the Knight field) acting
on the nuclei~\cite{DPJETP73}, the modulation of this field caused
by the modulated optical pumping is similar to the action of an RF
field~\cite{KalevichFTT80, KalevichFTT81, KalevichIzvNauk83,
KalevichFTT86, KalevichFTT95, KikkawaScience00, SalisPRL01,
SalisPRB01, EickhoffPRB02}. Optical detection of NMR under 
light-polarization modulation brings about a resonance cooling
of nuclear spins or, in other words, a cooling of nuclear spin 
system in the rotating frame~\cite{KalevichFTT80,KalevichFTT81}.
The effect was observed in a magnetic field oriented perpendicular 
to the excitation light beam (Voigt geometry). Resonance cooling 
results in appearance of Overhauser field oriented parallel or
antiparallel to the static magnetic field~\cite{KalevichFTT80, KalevichFTT81}. 

In our study we exploit another resonant effect connected with 
the modulation of the polarization of the optical pumping as described 
in Ref.~\onlinecite{CherbuninPRB11}. If the modulation frequency coincides 
with that of a nuclear-spin precession about the magnetic field, 
considerable nuclear-spin polarization perpendicular to the magnetic field 
appears in the system. This DNP component favorably increases 
the electron-spin polarization monitored in the experiment. As a result, 
additional maxima associated with the nuclear resonances appear in the 
magnetic field dependence of the electron spin polarization, i.e., 
in the Hanle curve. To accurately determine the positions of these nuclear 
resonances, we have measured the Hanle curves at different modulation 
frequencies varied from units of kilohertz up to of about $1$~MHz. An RF 
field synchronized with the optical modulation was applied to magnify 
the resonant effects.

\begin{figure*}[t]
\includegraphics[clip]{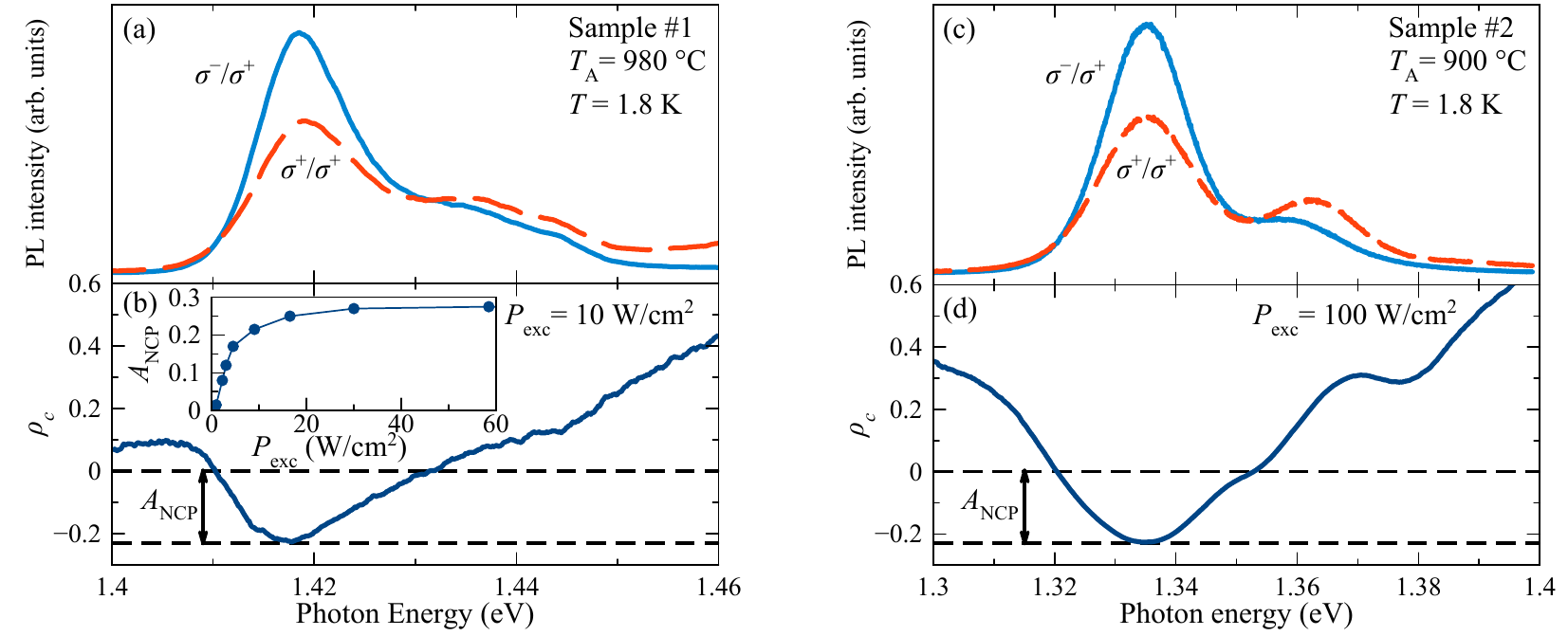}
\caption{(Color online) PL spectra (a) and (c) measured for
$\sigma^+$ excitation and co- and cross-polarized detection as well
as corresponding degree of circular polarization (b) and (d) for
samples~1 and 2. The definition of the amplitude of negative
circular polarization is illustrated by the arrow marked
$A_\mathrm{NCP}$ in (b) and (d). The inset in panel (b) shows
the power dependence of the NCP degree measured for sample~1. The
power dependence of $A_\mathrm{NCP}$ for sample~2 shows a similar
dependence (not shown here).}\label{fig:PL}
\end{figure*}

Identification of the different resonances was done by a theoretical
analysis based on modeling the quadrupole-Zeeman splitting of the
nuclear-spin states. Comparison of the calculated results with the
experimental data allows us to identify several resonances
corresponding to transitions between the states characterized by
spin projection $I_z = \pm 1/2$ for all the nuclei contained in the
QDs and by $I_z = \pm 3/2$ for the In and Ga nuclei. Besides,
resonances with transitions between the higher spin-projection
states, $I_z = \pm 5/2$,  for the In nuclei and between the states
having different projections, $I_z = \pm{1/2} \leftrightarrow I_z =
\pm 3/2$, for the In and Ga nuclei were identified. The large 
broadening of resonances is confirmed to be caused by the spread 
of strain within the QDs.

\section{Details of experiment}

We study a heterostructure containing 20 layers of self-assembled
(In,Ga)As QDs sandwiched between GaAs barriers. The barriers are
$\delta$-doped by donors with a concentration, which supplies every
dot with, on average, a single resident electron after the donor
ionization. The structure was grown by molecular-beam epitaxy on a
(001)-oriented GaAs substrate. Rapid thermal post-growth annealing
of the structure provides inter-diffusion of indium and gallium
atoms between the QDs and GaAs barriers so that the nominally pure
InAs QDs are partially enriched by the in-diffusion of Ga atoms
which increases the band gap, resulting in a blue shift of the
ground QD optical transition~\cite{LangbeinPRB04}. The annealing 
also causes a decrease of built-in strain in the QDs that is 
an important source of the quadrupole splitting of the nuclear-spin 
states. We study two pieces of this structure annealed at 
$T_\mathrm{A}=980$~\textdegree{}C (sample~1) and at 
$T_\mathrm{A}=900$~\textdegree{}C (sample~2).
The samples were placed in a cryostat with a superconducting magnet
such that the magnetic field could be applied perpendicular to the
structure growth axis (Voigt geometry) coinciding with the $[110]$
crystallographic axis. The experiments were performed at a sample
temperature $T=1.8$~K.

\begin{figure}[b]
\includegraphics[clip]{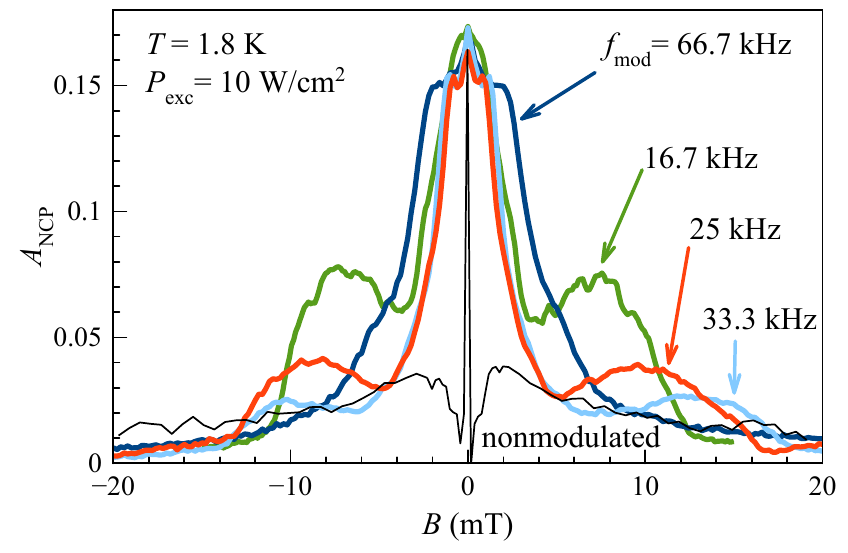}
\caption{(Color online) Hanle curves for sample 1 measured at
different frequencies of modulation of the excitation polarization.
Thin line shows the Hanle curve measured for nonmodulated
excitation.%
}\label{fig:Hanle}
\end{figure}

\begin{figure*}[t]
\includegraphics[clip]{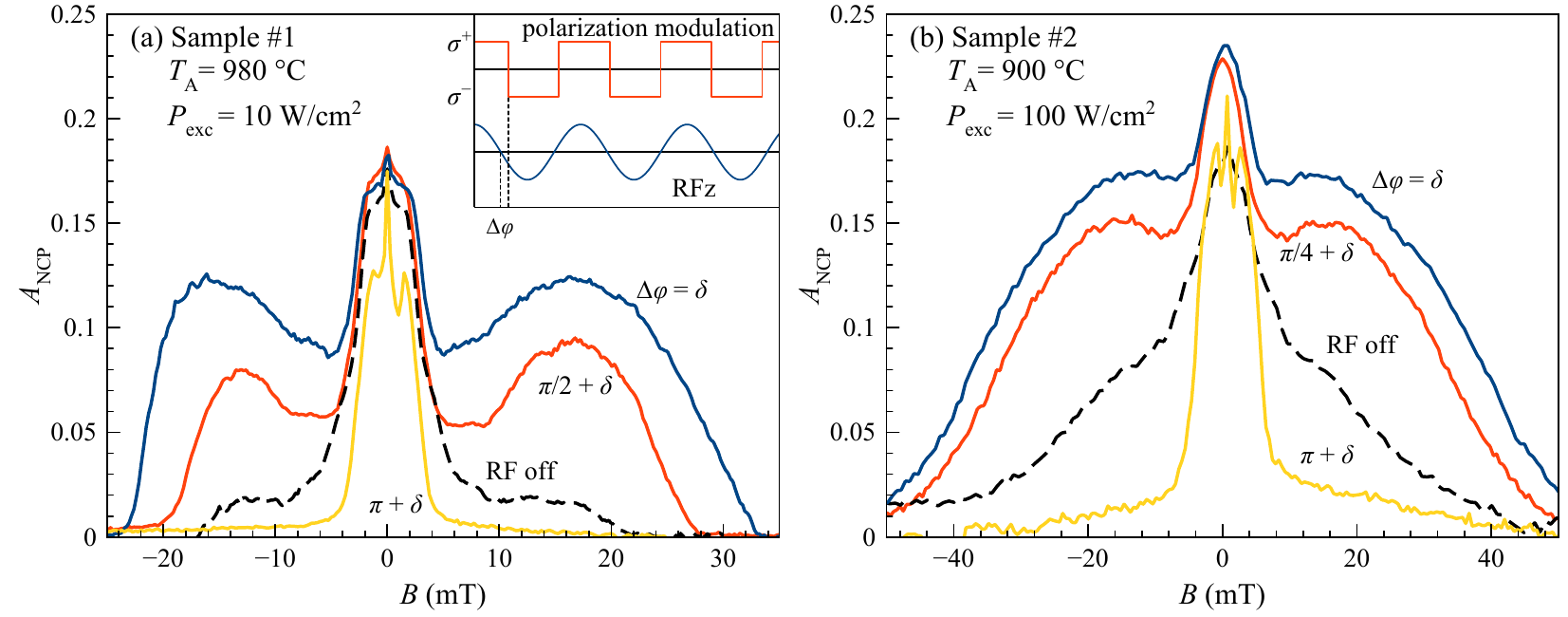}
\caption{(Color online) Joint action of RF~field application and
polarization modulation on the Hanle curves of sample~1 (a) and
sample~2 (b), measured at $f_\mathrm{mod} = 50$~kHz modulation
frequency. Resonant peculiarities in the Hanle curves can be either
amplified or suppressed (solid lines) as compared to the Hanle curve
without RF~field (dashed line), depending on the phase shift
$\Delta\varphi$ defined in the inset of (a).%
}\label{fig:Phase}
\end{figure*}

To detect DNP, we look over the circular polarization of the PL from
the QDs. To excite the QDs we use a continuous-wave Ti:sapphire
laser, which photon energy is tuned to the optical transition of the
wetting-layer exciton. The ground-state PL is dispersed by a half-metre 
spectrometer and is detected with a silicon avalanche photodiode.
The degree of circular polarization of the PL, $\rho_c = (I^{++} -
I^{+-})/(I^{++} + I^{+-})$, is measured using a photo elastic
modulator operated at a frequency of $50$~kHz and a two-channel
photon-counting system. Here, $I^{++}(I^{+-})$ is the PL intensity
for co- (cross-) polarization relative to that of excitation. In the
maximum of PL band of the QDs, the polarization is negative
(Fig.~\ref{fig:PL}) and reflects the mean spin polarization of the
resident electrons along the optical axis ($z$~axis), as has been
extensively discussed earlier~\cite{CortezPRL02, ShabaevPRB09,
IgnatievOS09}. Due to the interaction of the resident electrons with
the QD nuclei, the negative circular polarization (NCP) can be used
as a sensitive tool to monitor the nuclear spin
state~\cite{OultonPRL07, CherbuninPRB09, FlisinskiPRB10,
CherbuninPRB11}.

In our experiments, the helicity of the optical excitation was
periodically modulated between $\sigma^+$ and $\sigma^-$ by an
electro-optical modulator followed by a quarter-wave plate. In
addition, an RF field was applied by means of a small coil oriented
parallel to the optical $z$~axis. The RF-field frequency coinciding
with the frequency of optical modulation was fixed, while the Hanle
curves were measured by scanning the magnetic field.

\section{Experimental results and discussion}

\subsection{Hanle curves at optical excitation with modulated polarization. Dependence on modulation frequency}

The PL of the QDs at low excitation density shows a broad emission
contributed mostly by the ground state optical transition but also
the first excited state transition shows up on the high-energy
flank, see Fig.~\ref{fig:PL}. We focus here on the behavior observed
for the lowest transition. The corresponding emission line has the
half-width at half-maximum of about 10~meV, centered at 1.418~eV for
sample~1 and 1.337~eV for sample~2. The NCP is observed across
the whole emission line with a peak amplitude reaching 30\% at
optimal excitation density, which is about 30~W/cm$^2$ for
sample~1 and three times larger for sample~2. The analysis
shows~\cite{KuznetsovaPRB13} that the upper limit of NCP is
determined by the fraction of QDs containing a single resident
electron. Optimization of the excitation density allows one to fully
polarize the electrons. Note that the NCP is found to depend very 
sensitively on any residual magnetic field adding to the external 
field applied in the Voigt geometry. Therefore the residual magnetic 
fields including the Earth magnetic field were carefully compensated 
in our experiment by means of Helmholtz coils.

Magnetic-field dependencies of NCP were measured at the PL maximum
for different frequencies of polarization modulation of the
excitation. All Hanle curves presented in this paper were obtained
by scanning the magnetic field in one direction, defined here as
from negative to positive values. As an example, the Hanle curves
for sample~1 are shown in Fig.~\ref{fig:Hanle}. The Hanle curve
measured with nonmodulated polarization of the excitation (black
line) contains the so-called W-structure, which consists of a very
narrow central peak and two maxima separated from the peak by dips.
The appearance of the W-structure results from polarization of this
nuclear spins along the magnetic field~\cite{PagetPRB77}.

When the polarization of excitation is modulated, the Hanle curve
becomes strongly modified.
As seen from Fig.~\ref{fig:Hanle}, the central peak drastically
broaden and additional intense maxima appear. They move to higher
magnetic fields with increasing modulation frequency. The frequency
dependence indicates the resonant nature of the maxima. The rapid
decrease of their amplitude at higher frequencies does not allow,
however, to observe them in wide frequency range, which aggravates
their identification.

To increase the amplitude of the resonant peaks in the Hanle curves
and to extend the range of transverse magnetic fields, in which the
peaks can be observed, we applied to the sample an alternating RF
magnetic field of small amplitude along the optical axis. The
frequency of the alternating field was identical to that of the
modulation of the optical excitation. The phase difference,
$\Delta\varphi$, between the polarization and RF modulations was
varied.

The effect of the RF~field on the Hanle curve for sample~1 is
shown in Fig.~\ref{fig:Phase}. Obviously, the effect strongly depends
on $\Delta\varphi$~\cite{spuriousdelta}. At some optimal
$\Delta\varphi$, the wings of the Hanle curve are greatly increased
and the overall width of the curve is strongly
enlarged~\cite{hysteresis}. Phase inversion of the RF field relative
to the optimal case results in the opposite effect---the additional
wings almost disappear and only the central part of the Hanle curve
survives. For intermediate $\Delta\varphi$, an additional structure
appears in the Hanle curve, indicating asynchronous magnification or
suppression of the different resonances. This non-synchronicity
facilitates the separation of resonances. Similar experiments were
done for sample~2 where the Hanle curves are considerably wider
due to the stronger quadrupole interaction. The effect of the RF 
field for this sample is similar to the one for sample~1.

\begin{figure*}
\includegraphics[clip]{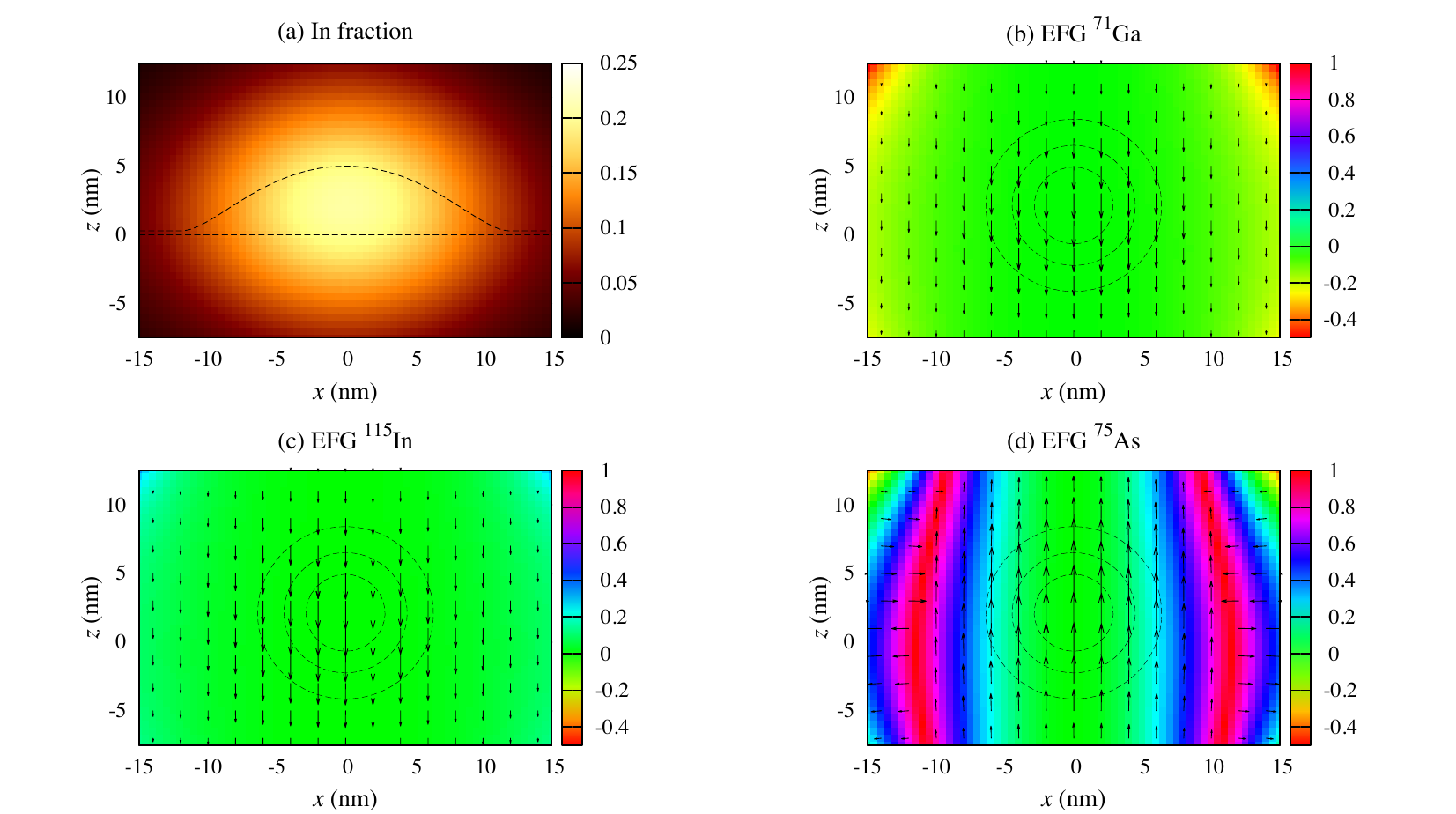}
\caption{(Color online) (a) Distribution of indium fraction
$x(\mathbf{r})$ in an annealed (In,Ga)As/GaAs QD. The boundary of
the unannealed QD is shown by the dashed line. (b)--(d)
Distributions of the parameter $\eta$ for different nuclei in the
annealed QD. The direction of the principal axis of the
electric-field gradient and its magnitude is shown by arrows. Dashed
lines show the isosurfaces of the electron density at levels of
$0.75$, $0.50$, and $0.25$ of the maximum from the middle to area,
respectively.%
}\label{fig:EFG}
\end{figure*}

\subsection{Identification of resonance peaks}

As seen from Fig.~\ref{fig:Phase}, the wings of the Hanle curve in
most cases are rather smooth due to the large resonance broadening,
whose origin we will discuss below. Therefore, to uniquely identify
the position of the NMR resonances for different nuclei isotopes, we
measured a set of Hanle curves at different modulation frequencies
$f_\mathrm{mod}$ and phase differences $\Delta\varphi$. The
amplitude of the RF field has been chosen to be relatively small
(fraction of millitesla) and has been optimized for the best separation of
the resonances. The experimental data were compared then with the
results of theoretical modeling of the Zeeman splittings of the
nuclear spin states in the QDs in presence of strain. The analysis
of the whole set of Hanle curves measured at different modulation
frequencies allowed us to identify the observed resonances, to
retrieve the nuclear Zeeman splittings, and to determine the nuclear
quadrupole splittings for the studied samples. The results of this
analysis are discussed below.

\subsubsection{Modeling of the quadrupole-Zeeman splitting of nuclear-spin states\label{sec:EFGModeling}}

An estimation of the crystal lattice deformation in QDs annealed at
different temperatures was given in
Ref.~\onlinecite{FlisinskiPRB10}. Here, we give insight into the
modeling of the strain-induced electric field gradient in the
annealed QDs. The results are summarized in Fig.~\ref{fig:EFG}.

The annealing of the (In,Ga)As/GaAs heterostructure initiates an
interdiffusion of In and Ga atoms followed by an increase
of the QD size and a decrease of the In dot content. We model this
process by solving a diffusion equation for the spatially
variable indium fraction, $x(\mathbf{r})$, in the In$_x$Ga$_{1-x}$As
compound (see Ref.~\onlinecite{PetrovPRB08} for details).
Figure~\ref{fig:EFG}(a) shows the cross-section of the resulting
distribution of $x(\mathbf{r})$ for the QD annealed at a temperature
above $900$~\textdegree{}C. It should be noted that the electron
localization volume increases with annealing due to both, the
increase of effective size of the dot and the decrease of the
potential well depth for the conduction-band
electron~\cite{PetrovPRB08}.

Additionally, the strain in the QD crystal lattice, arising from the
lattice mismatch of the QD and barrier materials, also changes with
annealing. For modeling the quadrupole-Zeeman splittings of the
nuclear sublevels in the QDs, it is important to account for the
component of biaxial strain, defined as $\varepsilon_B =
2\varepsilon_{zz} - \varepsilon_{xx} -
\varepsilon_{yy}$~\cite{BulutayPRB12}. This component relaxes from
the average value of $\varepsilon_B=0.13$ in the unannealed QD down
to $\varepsilon_B=0.01$ (sample 1) and $\varepsilon_B=0.03$
(sample 2) in the annealed QDs, as calculations show. These values
agree well with our experimental data as demonstrated below. The
calculations of $\varepsilon_B$ were made for a hat-shaped QD, which
we take as characteristic for the dots in the ensemble. Taking
advantage of the cylindrical symmetry of our model, the biaxial
component of the strain can be rewritten as $\varepsilon_B =
2\varepsilon_{zz} - \varepsilon_{rr} - \varepsilon_{\phi\phi}$. As
we are interested in ensemble-averaged values of the observables, the
strain-tensor components are calculated using continuum elasticity
theory using the transverse-isotropic-media
approximation~\cite{EvenAPL07,EvenPRB08}, which is valid in our case.
To obtain microscopic insight into an individual QD, a more
complicated atomistic approach should be taken~\cite{BulutayPRB12}.

The crystal lattice deformation is enhanced within the QD and causes
there the appearance of an electric field gradient (EFG) at nuclear
sites. The main origin of the EFG is the strain due to lattice
mismatch. There is also a crystal field gradient due to the
statistical population of anion sub-lattice sites with In and Ga
atoms. This statistics concerns a fraction of the As nuclei and
gives rise to such a strong quadrupole splitting that the
corresponding resonances can be observed only in very high magnetic
fields, which are available in our experiments.

The EFG relates to the elastic strain by a fourth-order
tensor~\cite{ShulmanPR57}, i.e.,
\begin{equation}
    V_{ij} = S_{ijkm} \varepsilon_{km},
\label{eq:EFG}
\end{equation}
where summation over iterated subscripts is assumed. For
$\mathrm{A_{III}B_V}$ compounds, there are only three nonzero
components of the $S$-tensor and only two of them are
independent~\cite{SundforsPRB74}, $S_{11}$, $S_{12} =
\tfrac{1}{2}S_{11}$, and $S_{44}$, written in the Voigt notation.

The goal of our modeling is to obtain the principle components of
the strain-induced EFG, $V_{11}$, $V_{22}$, and $V_{33}$. In the
first step, we calculate all nonzero components of $V_{ij}$ using
Eq.~\eqref{eq:EFG} and, in the second step, we diagonalize the
obtained matrix $V$. By doing this, we find the spatial dependencies
of the principal component of the EFG, $V_{33}$, and the asymmetry
parameter, $\eta = (V_{11}-V_{22})/V_{33}$, for all the nuclei in
the dot. The results of these calculations are shown in
Figs.~\ref{fig:EFG}(b)--\ref{fig:EFG}(d) for the $^{71}$Ga,
$^{115}$In, and $^{75}$As nuclei, respectively. As one can see from
the figures, the principal component of the EFG tensor is directed
along the $z$-growth axis of the structure for all the nuclei.
Additionally, the EFG direction at the As nuclei is opposite to that
for the other nuclei because of the different signs of $S_{11}$ and
$S_{44}$ for anions and cations (see Ref.~\onlinecite{SundforsPRB74}
for details). Large variations of the parameter $\eta$ are found for
the As nuclei located in the barrier to the left and right of the
QD. Inside the dot where the ground-state electron density is large,
this variation is negligibly small. To illustrate that, we show in
Figs.~\ref{fig:EFG}(b)--\ref{fig:EFG}(d) the isosurfaces of the
electron density distribution, calculated in the effective-mass
approximation~\cite{PetrovPRB08}.

\begin{figure*}[t]
\includegraphics[clip]{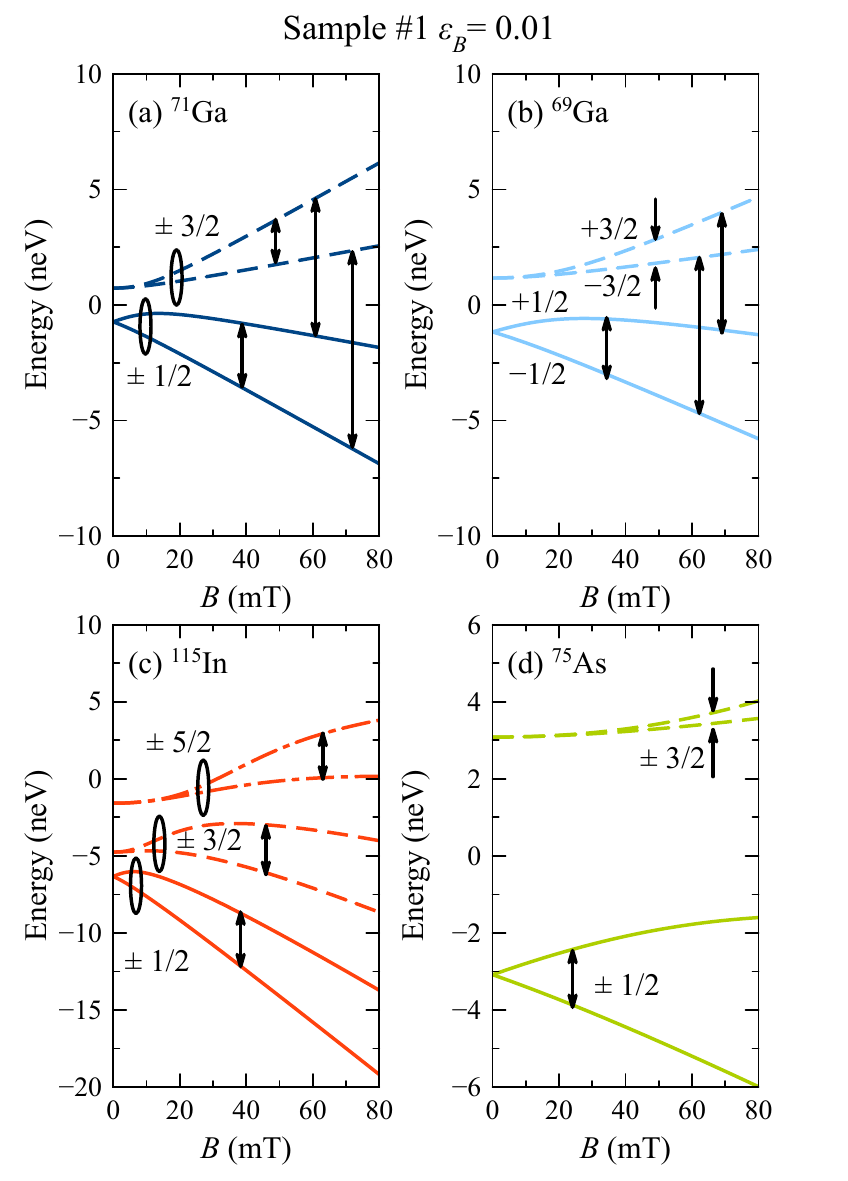}
\includegraphics[clip]{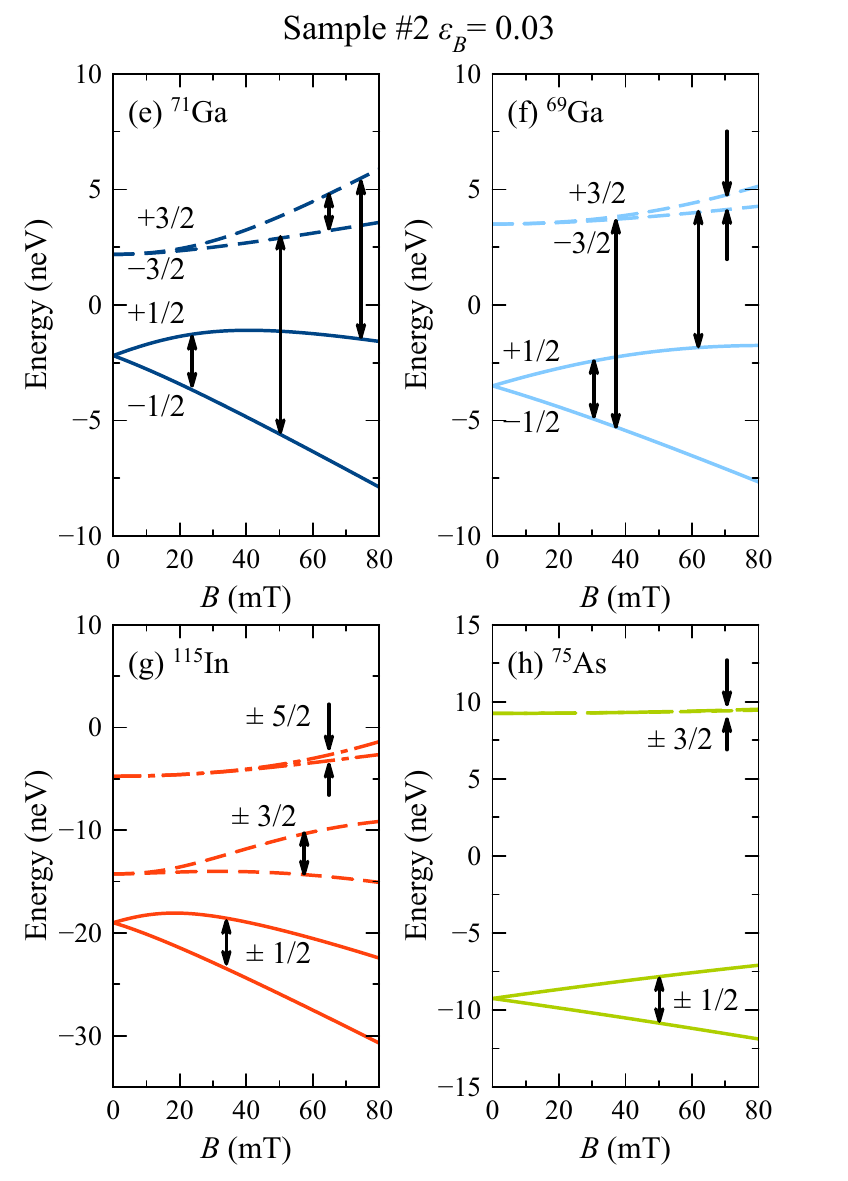}
\caption{(Color online) Energies of nuclear-spin sub-levels
calculated for the isotopes $^{71}$Ga, $^{69}$Ga, $^{115}$In, and
$^{75}$As as functions of magnetic field applied along the $x$~axis
for sample~1 (a) and sample~2 (b). Quadrupole constants and
gyromagnetic ratios for the isotope $^{115}$In are close to those
for the $^{113}$In isotope~\cite{PyykkoMolPhys08, HarrisPApCh01},
therefore the energy dependences for $^{113}$In and $^{115}$In
almost coincide. Arrows show the nuclear-spin transitions discussed
in the text. Note, that they are labeled for simplicity by pure spin
states $\ket{\pm1/2}$ etc. even though the states are mixed by the magnetic field.%
}\label{fig:Fitting}
\end{figure*}

\begin{figure*}
\includegraphics[width=.95\columnwidth, clip]{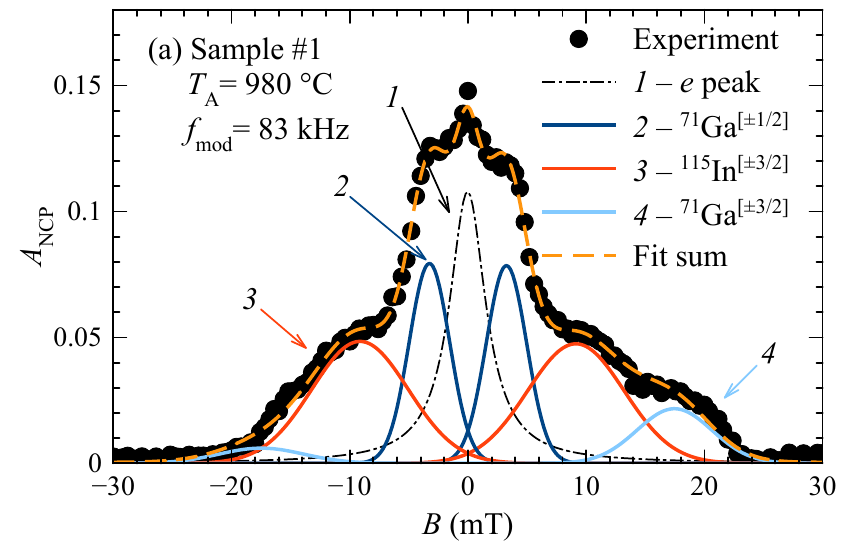}
\hspace{.1\columnwidth}
\includegraphics[width=.95\columnwidth, clip]{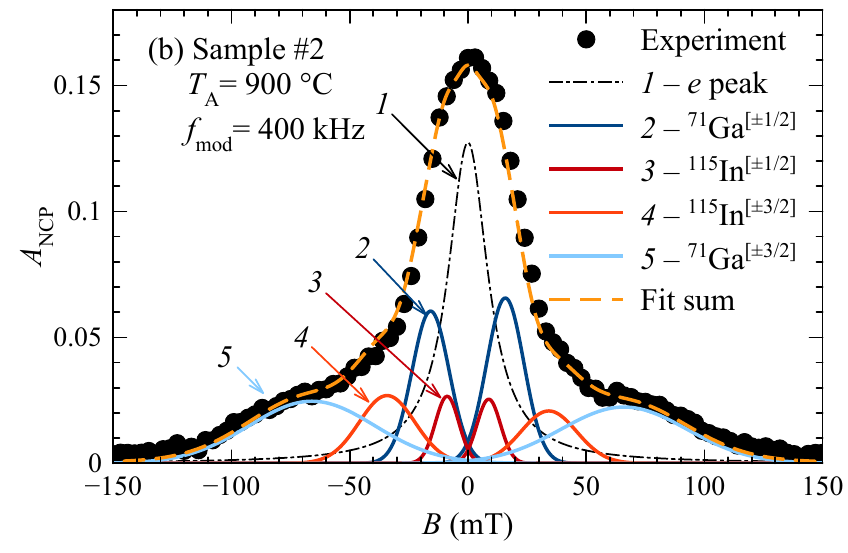}
\caption{(Color online) Deconvolution of the experimentally measured
Hanle curves into resonances corresponding to the transitions
$+1/2\leftrightarrow-1/2$ and $+3/2\leftrightarrow-3/2$ in the
nuclear-spin system. (a) Sample~1, modulation frequency
$f_\mathrm{mod} = 83$~kHz; (b) sample~2, $f_\mathrm{mod} =
400$~kHz. Points represent the experimental data. Dash-dotted lines
are Lorentzians corresponding to the $e$~peak. Solid lines are the
Gaussians modeling the resonances, and dashed lines represent the results of fits.%
}\label{fig:HanleFit}
\end{figure*}

At zero magnetic field, the nuclear-spin states are split by the
quadrupole interaction into doublets
$\pm{1/2},\pm{3/2},\ldots,\pm{I/2}$ for nuclei with spin $I > 1/2$,
possessing a non-zero quadrupole moment $Q$. The magnitude of the
splitting, expressed in terms of the frequency $\nu_Q$, is
determined by~\cite{Slichter92}
\begin{equation} \label{eq:Eq1}
    h\nu_Q=\frac{3eQ V_{33}}{2I(2I-1)}.
\end{equation}
Since the principal component of EFG is aligned along the $z$~axis,
we set
\begin{equation} \label{eq:Eq2}
    V_{33} = \tfrac12 S_{11}\varepsilon_B.
\end{equation}

As the asymmetry parameter $\eta$ of the quadrupole interaction is
neglected, the Hamiltonian describing the splitting of the
nuclear-spin states is~\cite{CommentQZHam}
\begin{equation}
    \hat{\mathcal{H}} = - \hbar\gamma_I B \hat{I}_x + \frac{h\nu_Q}{2}
    \left[\hat{I}_z^2-\frac{I(I+1)}{3}\right],
\label{eq:Ham}
\end{equation}
where $\gamma_I$ is the gyromagnetic ratio. The first term of the
Hamiltonian describes the Zeeman interaction and the second term is
the quadrupole interaction. The matrix of this Hamiltonian using the
basis states $\ket{\pm{1/2}}, \ket{\pm{3/2}}, \ldots,
\ket{\pm{I/2}}$ is diagonalized to obtain the energies of the
nuclear spins under the influence of the Zeeman and quadrupole
interactions. For the calculations, we use the values of $S_{11}$,
$Q$, and $\gamma_I$ collected in Table~\ref{tab:NucParam}.

\begin{table}[b]
\caption{Summary of nuclear parameters. The values of $S_{11}$, $Q$,
and $\gamma$ are taken from Refs. \onlinecite{SundforsPRB74},
\onlinecite{PyykkoMolPhys08}, and \onlinecite{HarrisPApCh01},
respectively. }\label{tab:NucParam} \center
\begin{tabular}{l c c c c}
\hline
\hline
Isotope & $I$ & $\gamma_I$ (rad s$^{-1}$ T$^{-1}$) & $Q$ (mbar) & $S_{11}$ (statC cm$^{-3}$)\\
\hline
$^{69}$Ga & 3/2 & $6.439\times10^7$ & 171 & $9.1\times10^{15}$ \\
$^{71}$Ga & 3/2 & $8.181\times10^7$ & 107 & $9.1\times10^{15}$ \\
$^{75}$As & 3/2 & $4.596\times10^7$ & 314 & $1.31\times10^{16}$ \\
$^{113}$In & 9/2 & $5.885\times10^7$ & 759 & $1.67\times10^{16}$ \\
$^{115}$In & 9/2 & $5.897\times10^7$ & 770 & $1.67\times10^{16}$ \\
\hline
\hline
\end{tabular}
\end{table}

The spin splittings calculated for the Ga, As, and In nuclei in the
QDs under study are shown in Fig.~\ref{fig:Fitting}. One can see
that the resonances for transitions between the split-off states
$\ket{\pm{3/2}}, \ldots, \ket{\pm{I/2}}$ are observed at larger
magnetic fields than those for the $\ket{\pm{1/2}}$ states. Besides,
the resonances at a given frequency for the split-off states are
shifted to larger magnetic fields in sample 2 relative to those in
sample~1 due to the larger $\nu_Q$.

\begin{figure*}
\includegraphics[clip]{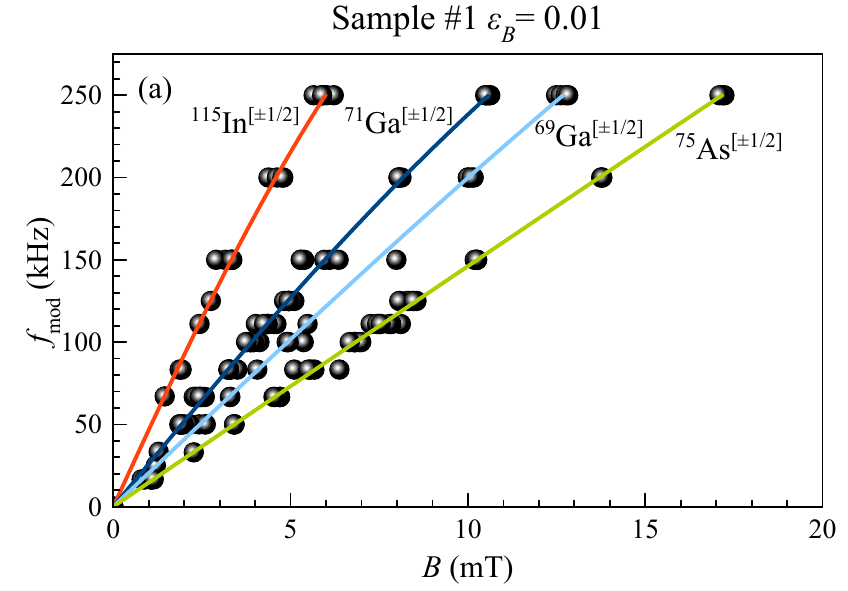}
\includegraphics[clip]{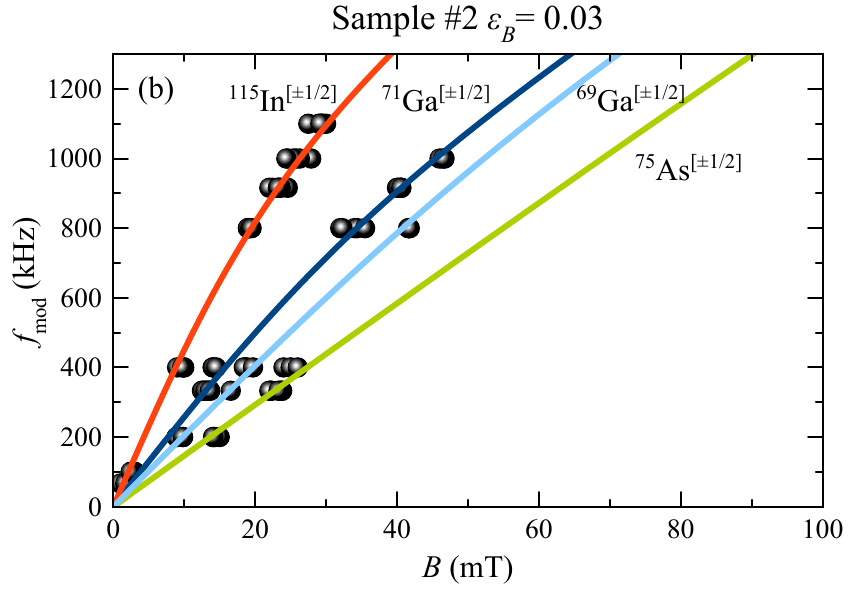}\\
\includegraphics[clip]{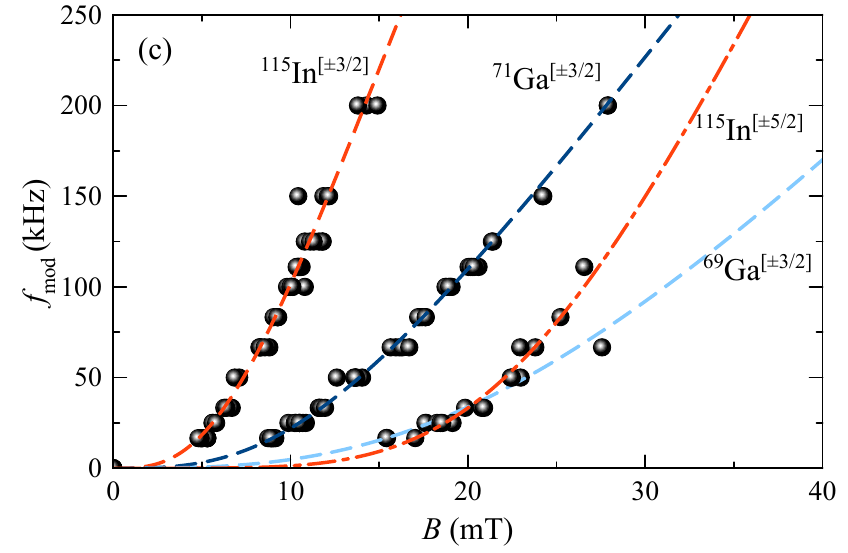}
\includegraphics[clip]{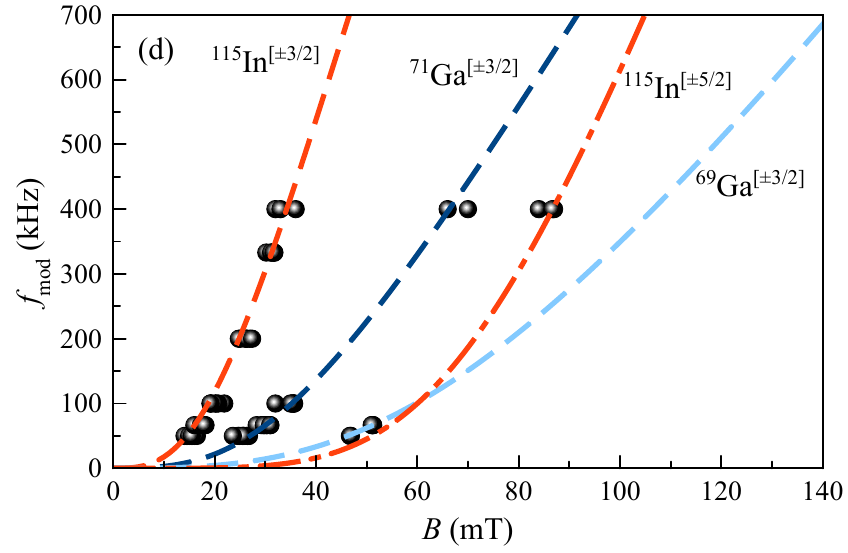}\\
\includegraphics[clip]{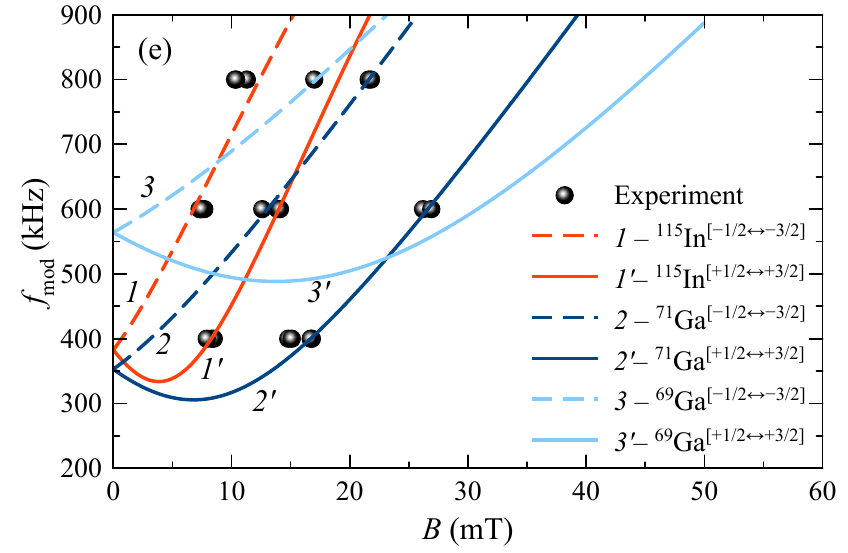}
\includegraphics[clip]{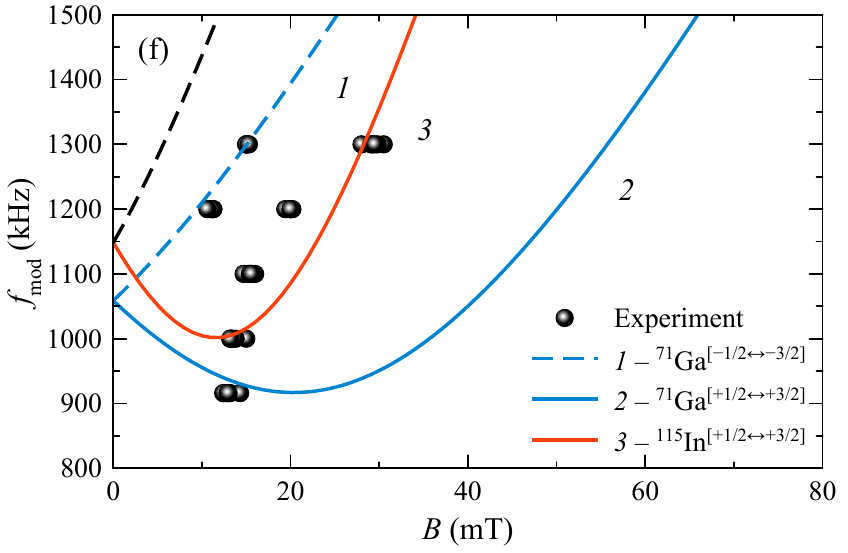}
\caption{(Color online) Magnetic-field dependencies of the
nuclear-spin transition frequencies $+1/2 \leftrightarrow -1/2$
(a)--(b), $+3/2 \leftrightarrow -3/2$ and $+5/2 \leftrightarrow
-5/2$ (for In nuclei only) (c)--(d), and $\pm1/2 \leftrightarrow
\pm3/2$ (e)--(f). Points are the positions of the resonant peaks
obtained from the experiment for different phase shifts
$\Delta\varphi$. Lines are the calculations.}
 \label{fig:ResonancePositions}
\end{figure*}

\subsubsection{Fitting of the resonances}

The shape of the resonances contributing to the Hanle curves is
determined by their physics origin. Under action of
optical pumping with non modulated polarization,
DNP appears aligned parallel to the external magnetic field. The
alternating magnetic field from the radio-frequency excitation close
to the resonant frequency of nuclear spin transitions, applied along
the optical axis, tilts this nuclear-spin magnetization. When
scanning the external field around the resonance, the tilting angle
changes from $0$ to $\pi$ taking on the value $\pi/2$ at the
resonance. Consequently, the sign of the projection of the
nuclear-spin magnetization onto the axis of external magnetic field
changes while scanning through the resonance. This results in a
dispersion like singularity in the Hanle curve. Such resonant
profiles were previously observed in different systems and discussed
in literature~\cite{Paget-OO, SalisPRL01, SalisPRB01,
EickhoffPRB02}.

In contrast to non modulated
excitation, optical pumping with circular polarization modulated at
a frequency close to a resonance in a QD ensemble leads to a
nuclear-spin magnetization aligned \emph{perpendicular} to the
external magnetic field. This DNP component is created by nuclear
spins, which coherently precess in the external magnetic field.
Synchronization of the precession of the DNP field with the
modulated Knight field maintains the electron-spin polarization. The
latter results in positive peaks on the Hanle curve.
These resonances can be observed outside the electronic peak
($e$~peak) in the Hanle curve given by pure electronic
polarization~\cite{CherbuninPRB11}. Analysis of our experimental
data shows that all the resonances must be modeled by positive peaks
rather than the dispersion like curves for which we used Gaussians.

The resonances corresponding to the $+1/2\leftrightarrow-1/2$
transitions overlap so that their phenomenological analysis does not
allow us to derive definite conclusions about their nature. To
simplify the fit of experimental data, we also used Gaussians for
their modeling~\cite{Note-Yugova}. So, it is assumed that each Hanle
curve is a superposition of Gaussian peaks centered near the NMR
positions calculated for each type of nuclei. Besides, the Hanle
curve is contributed by an electronic peak ($e$~peak) observed in
absence of DNP. The shape of this peak is determined experimentally
by measuring the Hanle curve using an amplitude modulation of the
excitation with a large ratio of dark to bright intervals. As shown
in Ref.~\onlinecite{CherbuninPRB11}, DNP is strongly suppressed
under such  experimental conditions.

The calculation of splittings described above allows us to determine
the magnetic-field positions of nuclear-spin resonances at each
particular modulation frequency and, thus, to fit the Hanle curves.
The fitting parameters are the width and the amplitude of the
Gaussians. The magnetic-field position of the resonances is taken,
at the first step, from the modeling (see Fig.~\ref{fig:Fitting})
and then slightly varied to obtain the best fit to the experiment.
For more precise determination of the resonance positions, a set of
Hanle curves measured with different phase shifts $\Delta\varphi$
between the polarization modulation and the RF field is analyzed.
The resonance positions are found to be close to those obtained from
the modeling. However, the widths of the resonances are much larger
than those typically observed in standard NMR. The origin of this
broadening of the resonances will be discussed in the next section.

Examples of deconvolutions of the Hanle curves into Gaussian-like
resonances are given in Fig.~\ref{fig:HanleFit}. The figure shows
the Hanle curves measured at moderate modulation frequencies when
only the transitions  $+1/2\leftrightarrow-1/2$  and
$+3/2\leftrightarrow-3/2$ contribute to the curves. As one can see,
the strongly broadened resonances overlap. The central part of the
Hanle curve for sample~1 can be well modeled by the sole
contribution of the transitions $+1/2\leftrightarrow-1/2$ of
$^{71}$Ga isotope. In the Hanle curve of sample~2, contributions
of transitions in the In nuclei can be also resolved. The wide part
of Hanle curves is given by the transitions
$+3/2\leftrightarrow-3/2$ of the In and Ga nuclei. The relative
amplitude of these resonances is found to be very sensitive to the
experimental conditions, in particular, to the phase shift
$\Delta\varphi$ and does not reflect the content of In and Ga nuclei
in the QDs. For example, although the In content is larger in
sample~2, the respective resonance $+3/2\leftrightarrow-3/2$ of
the In nuclei gives rise to a stronger peak in the Hanle curve of
sample~1 for the experimental conditions used for recording the
data in Fig.~\ref{fig:HanleFit}.

A comparison of the resonance positions obtained from the experiment
with those obtained from the calculation of the Zeeman splittings of
the nuclear-spin states is shown in
Fig.~\ref{fig:ResonancePositions}.
Figures~\ref{fig:ResonancePositions}(a)--\ref{fig:ResonancePositions}(d)
demonstrate the data for the observed transitions
$+1/2\leftrightarrow-1/2$, $+3/2\leftrightarrow-3/2$, and
$+5/2\leftrightarrow-5/2$. Variations of the phase shift
$\Delta\varphi$ between the polarization modulation and the RF field
slightly change the resonance positions. This is reflected in some
horizontal spread of the experimental data in the figures.
Figures~\ref{fig:ResonancePositions}(e) and
\ref{fig:ResonancePositions}(f) show similar data for the
transitions $+1/2\leftrightarrow+3/2$  and
$-1/2\leftrightarrow-3/2$. The resonance frequencies for these
transitions can be determined only with less accuracy. However, the
Hanle curve cannot be successfully described if these resonances are
neglected.

The results obtained show that, in spite of the large broadening of
the resonances, most of them can be identified and their behavior
can be described by a simple model containing only one free
parameter, $\varepsilon_B$. The values of this parameter obtained
from experiment for the samples under study are very close to those
calculated (see Sec.~\ref{sec:EFGModeling}).

\begin{figure}[t]
\includegraphics[clip]{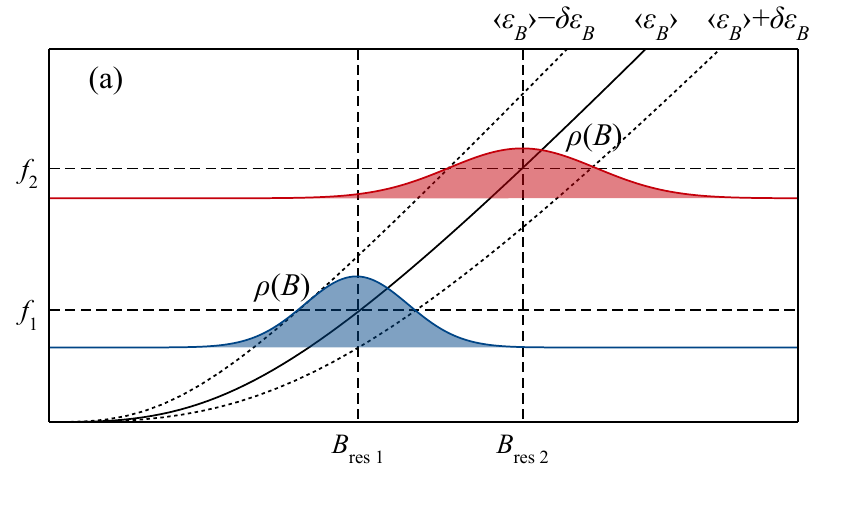}\\
\includegraphics[width=\columnwidth,clip]{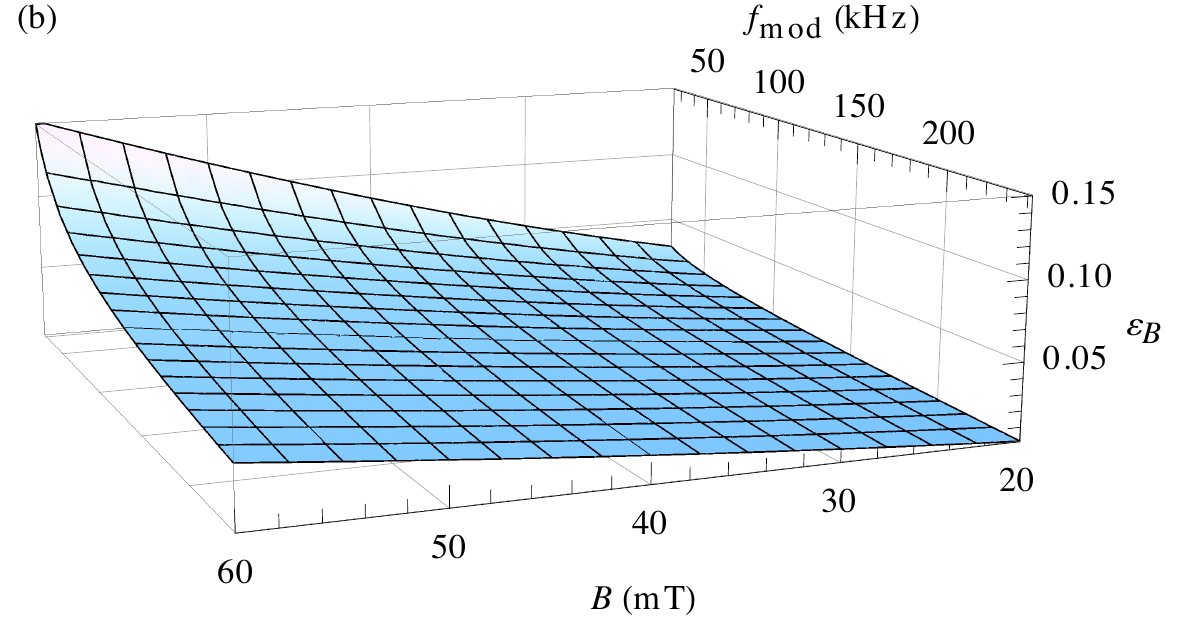}
\caption{(Color online) (a) Zeeman splittings of the split-off
states $\ket{\pm3/2}$, experiencing some spread due to the
inhomogeneity of the biaxial strain  (solid and dotted lines).
Because of this spread, the resonances $\rho(B)$ broaden and the
broadening increases with $f_\mathrm{mod}$ (for details see the
text). (b) Relation between the biaxial strain and the
magnetic-field position of the resonance at different frequencies of
the transitions $+3/2\leftrightarrow-3/2$ in the $^{71}$Ga nucleus.
}\label{fig:3DStatistics}
\end{figure}

\begin{figure}
\includegraphics[clip]{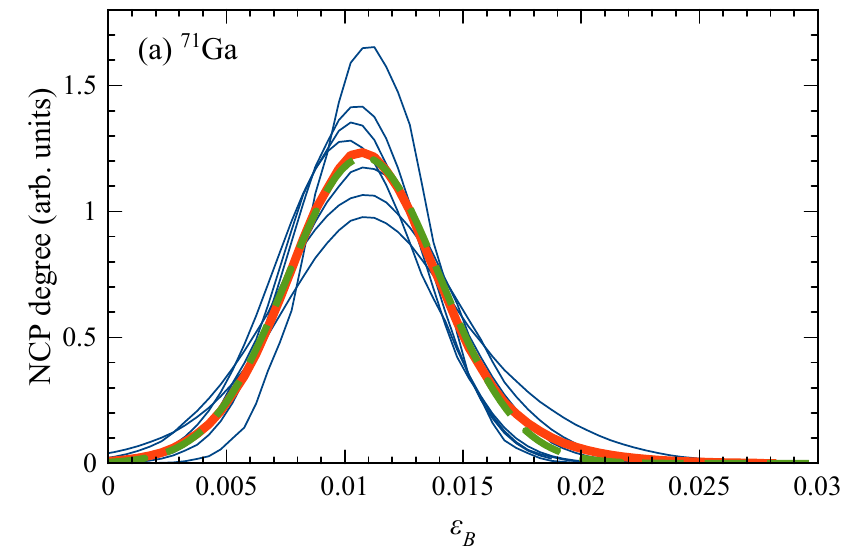}\\
\includegraphics[clip]{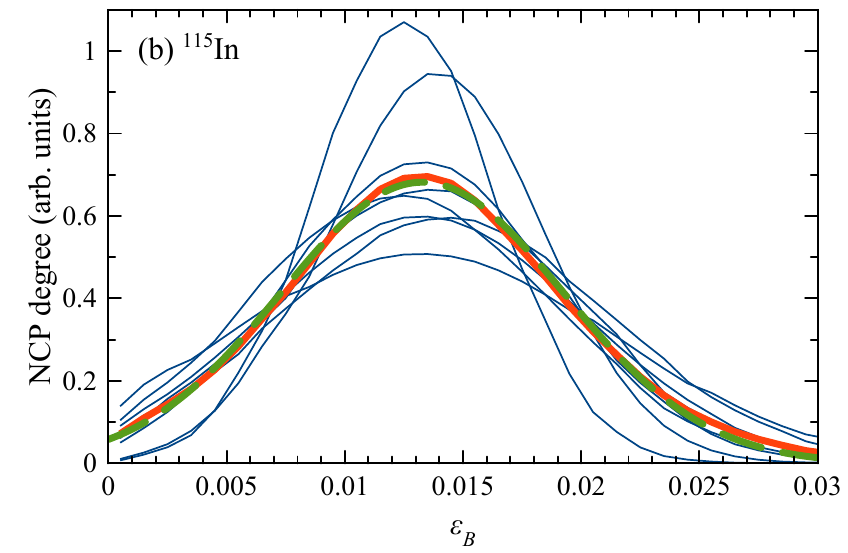}
\caption{(Color online) Resonance profiles as functions of biaxial
strain obtained from experimental data measured in the range from
$f_\mathrm{mod}=10$ to $250$~kHz for sample~1. All resonances are
normalized to the same integral area. The resonances corresponding
to the transitions $+3/2\leftrightarrow-3/2$ for $^{71}$Ga are shown
in panel (a) and for $^{115}$In in panel (b). The thick solid lines
are the averaged profiles and the dashed lines are Gaussian fits to
them: $\rho(\varepsilon_B) = \rho_0 \exp[-(\varepsilon_B
-\braket{\varepsilon_B})^2/\sigma^2]$. The fitting parameters are
$\braket{\varepsilon_B} = 0.011$ and $\sigma = 0.0048$ for
$^{71}$Ga,
$\braket{\varepsilon_B} = 0.013$ and $\sigma = 0.0085$ for $^{115}$In.\\
}\label{fig:Statistics}
\end{figure}

\subsubsection{Width of resonances and its relation to strain}

The fitting of the Hanle curves by sums of resonant peaks (see
Figs.~\ref{fig:HanleFit} and \ref{fig:ResonancePositions}) has
revealed the large widths of the resonances, which are considerably
broader than those observed for quantum wells~\cite{SalisPRL01,
SalisPRB01, EickhoffPRB02}. Particularly, large widths are observed
for the peaks corresponding to the transitions
$+3/2\leftrightarrow-3/2$ and $+5/2\leftrightarrow-5/2$. These
widths are about an order of magnitude larger than those caused by
random dipole-dipole fields (fraction of mT) or by the Knight field
(of about $1$~mT in sample~1, see
Ref.~[\onlinecite{KuznetsovaPRB13}]). Variation of these fields has
been considered in Ref.~[\onlinecite{UrbaszekRMP13}] as the origin of
broadening the NMR resonances in single QDs.

We assume that a more broader effect on the resonance broadening
originates from the spread of strain causing a spread of the
quadrupole splittings of the nuclear-spin states. The spread can be
present in each individual QD~[\cite{ChekhovichNatNano12}] and in the
QD ensemble as a whole. The Zeeman splitting of the nuclear-spin
states strongly depends on the biaxial strain. As one can see from
Figs.~\ref{fig:ResonancePositions}(a)--\ref{fig:ResonancePositions}(e),
the splitting in sample~2 with large strain is smaller than that
in sample~1 for the same magnetic field. At a fixed resonance
frequency, each particular magnitude of $\varepsilon_B$ corresponds
to a unique magnitude of magnetic field, $B_\mathrm{res}$, at which
the resonance is observed. Correspondingly, the spread of strain
gives rise to a spread of resonant magnetic fields as illustrated in
Fig.~\ref{fig:3DStatistics}(a). There examples of resonance
profiles, $\rho(B)$, are shown, which describe the probability of
NMR as a function of magnetic field at a given frequency. As seen,
the profiles are shifted to higher magnetic fields and become
broader with increasing frequency which qualitatively explains the
experimental observations.

The shape of the resonance profiles $\rho(B)$ should be
unambiguously related to the function $\rho'(\varepsilon_B)$
describing the spread of biaxial strain in the structure. To
determine the relation between the functions $\rho(B)$ and
$\rho'(\varepsilon_B$), we have calculated the magnetic-field
dependence of the splitting of the nuclear-spin states $\pm{3/2}$
for the Ga and In nuclei at different values of $\varepsilon_B$.
Results of these calculations for the $^{71}$Ga isotope are shown in
Fig.~\ref{fig:3DStatistics}(b). As one can see, there is a
monotonic, almost linear, dependence, $B_\mathrm{res} \approx a_f
\varepsilon_B$, for modulation frequencies $f_\mathrm{mod}$ fixed in
the range used in experiment. This allows one to interconnect the
two probability distributions:
\begin{equation}
\rho(B)=\rho'(\varepsilon_B)d\varepsilon_B/dB,
\end{equation}
where the derivative $d\varepsilon_B/dB$ can be easily calculated.
Figure ~\ref{fig:3DStatistics}(b) shows that the derivative weakly
depends on $B$ and is a function of $f_\mathrm{mod}$.

The resonance peaks, $\rho({B_\mathrm{res}})$, obtained from the
experiments and modeled by Gaussians, can be treated except for some
scaling factor, as the probability distribution $\rho(B)$ of NMR at
a given frequency. Using the equation given above one can transform
$\rho({B_\mathrm{res}})$ into $\rho'(\varepsilon_B)$. Because the
spread of strain does not depend on modulation frequency, the
resonant peaks measured at different frequencies and transformed
into functions of $\varepsilon_B$ should be very similar to each
other, if the broadening is actually due to this spread.

The results of such a processing of our experimental data for the
transitions $+3/2 \leftrightarrow -3/2$ of the $^{71}$Ga and
$^{115}$In nuclei are shown in Fig.~\ref{fig:Statistics}.
As seen, variation of the modulation frequency over a wide range
from $10$ to $250$~kHz weakly affects the positions and the widths
of the resonance profiles $\rho'(\varepsilon_B)$. Note, that the
width of $\rho({B_\mathrm{res}})$ increases several times with the
modulation-frequency
The relatively small variations of the parameters of the
experimentally determined resonance profiles shown in
Fig.~\ref{fig:Statistics} are probably caused by experimental errors
and some uncertainty in the fitting procedure. There is also some
difference in the averaged positions and widths of the
$\rho'(\varepsilon_B)$ for the nuclei $^{71}$Ga and $^{115}$In.

\begin{figure}[t]
\includegraphics[clip]{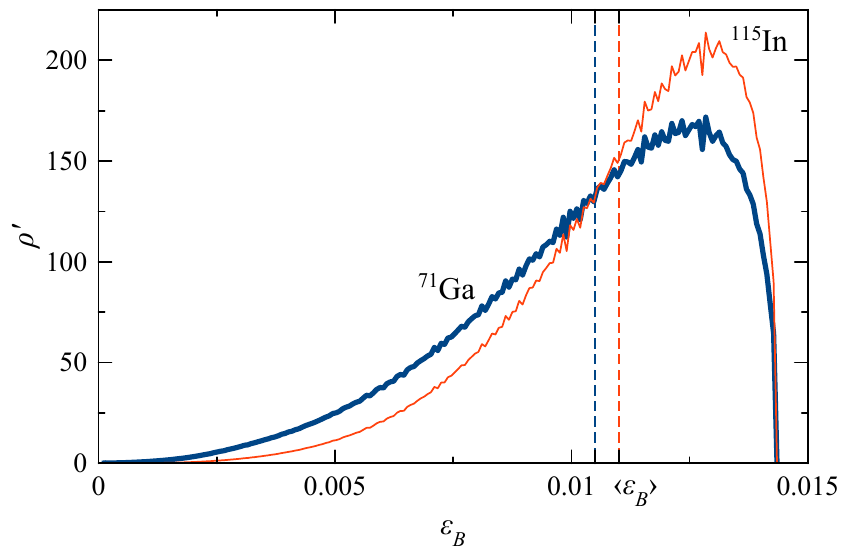}
\caption{(Color online) Statistics of the resonances $+3/2
\leftrightarrow -3/2$ in the $^{71}$Ga (thick blue line) and
$^{115}$In (thin red line) nuclei calculated in the framework of the
model described in Sec.~\ref{sec:EFGModeling} for the sample~1
annealed at 980 \textdegree{}C. Dashed lines indicate the mean
values $\braket{\varepsilon_B} = 0.0105$ for $^{71}$Ga and
$\braket{\varepsilon_B} = 0.0111$ for $^{115}$In. }
\label{fig:rho_th}
\end{figure}

To model the function $\rho'(\varepsilon_B)$ in a single QD, we
perform a simple calculation based on the model described in
Sec.~\ref{sec:EFGModeling}. Using the distribution of the In
fraction, $x(\mathbf{r})$, and the electron density distribution,
$|\psi_e(\mathbf{r})|^2$, one can estimate the distribution of the
principal component of the EFG tensor by
\begin{equation}
\rho'(\varepsilon_B) = \iiint_{\mathbb{R}^3} x(\mathbf{r})
|\psi_e(\mathbf{r})|^2
\delta[V_{33}(\mathbf{r}) - \tfrac{1}{2} S_{11} \varepsilon_B]
d\mathbf{r}.
\label{eq:rho_th}
\end{equation}
To obtain equivalent distributions for the other isotopes one should
replace $x(\mathbf{r})$ by $\left(1-x(\mathbf{r})\right)$ (for Ga)
and by $1$ (for As) in Eq.~\eqref{eq:rho_th}.

The resulting distributions $\rho'(\varepsilon_B)$ for the In and Ga
nuclei are shown in Fig.~\ref{fig:rho_th}. As one can see, the
distributions are relatively wide, which supports the assumption of
strain non-homogeneity within the QD as primary source of the
resonance broadening shown in Fig.~\ref{fig:Statistics}. There is a
strong asymmetry of the distribution with an abrupt right edge,
which corresponds to the maximal value of $\varepsilon_B$ in the
middle of the QD. The experimentally obtained distribution shown in
Fig.~\ref{fig:Statistics} is more symmetric, which is possibly due
to smoothing of the distribution in the QD ensemble. There is also
some difference in the values $\braket{\varepsilon_B}$ for $^{71}$Ga
and $^{115}$In averaged over the corresponding statistics. The In
content is maximal in the QD middle where the EFG is also maximal so
that the distribution for the In nuclei is shifted to larger values
of $\varepsilon_B$ relative to those for the Ga nuclei. This result
correlates well with the experimental observations, see
Fig.~\ref{fig:Statistics}.

The broadening of the resonance peaks for the transitions
$+1/2\leftrightarrow-1/2$ requires some further consideration. This
broadening is considerably smaller than that for the transitions
$+3/2\leftrightarrow-3/2$. However, it is still much larger than the
one typically observed in solid state NMR~\cite{Abragam}. To
understand the possible origin of this broadening we should consider
the Zeeman splitting of the states $\ket{\pm{1/2}}$ in presence of
quadrupole splitting. When the magnetic field is small and mostly
perpendicular to the principal axis of the EFG, the splitting is
given by~\cite{PoundPR50}
\begin{equation} \label{eq:Splitting}
    \Delta{E}\approx 2E_Z\left[1-\frac{3}{16}\left({{\frac{\Delta{E_Z}}{\Delta{E_Q}}}}\right)^2\right],
\end{equation}
where $\Delta{E_Z}$ is the Zeeman splitting of the states
$\ket{\pm{1/2}}$ in absence of strain and $\Delta{E_Q}$ is the
quadrupole splitting of the nuclear-spin states into doublets
corresponding to $\ket{\pm{1/2}}$ and $\ket{\pm{3/2}}$. Clearly,
when $\Delta{E_Z}\ll\Delta{E_Q}$, the splitting $\Delta{E}$ is
approximately twice larger than the Zeeman splitting and almost
independent of the deformation. This means that the spread of
$\varepsilon_B$ cannot be responsible for the observed broadening.
We have to assume that the broadening is caused by a deviation of
the principle axis of the EFG from the direction orthogonal to the
magnetic-field. One possible reason for that is an asymmetry of the
QDs in the ensemble due to the statistical nature of the assembly of
atoms in the QDs during the growth process. Another reason might be
the inclination of the deformation axis at peripheral parts of the
QDs~\cite{BulutayPRB12}. Note that even a relatively small
inclination of the axis may cause a remarkable shift of the
resonance position. Indeed, when the EFG axis is parallel to the
magnetic field (inclination is $\pi/2$), the splitting of the
$\ket{\pm{1/2}}$ states becomes twice smaller and corresponds to the
ordinary Zeeman splitting, $\Delta{E} = \Delta{E_Z}$. Therefore the
spread of inclinations of the EFG axis in the QD ensemble can be
responsible for the observed broadening of the resonances $+1/2
\leftrightarrow -1/2$.

\section{Conclusion}

To conclude, the obtained results demonstrate that resonant optical
pumping of the electron-nuclear spin system in QDs subject to
transverse magnetic fields is an efficient tool for studying
transition between nuclear spin states split by a magnetic field.
Using this method, we managed to experimentally detect a number of
resonances of the In, Ga, and As nuclei in an inhomogeneous QD
ensemble, namely those corresponding to the transitions between the
$+1/2 \leftrightarrow -1/2$, $+3/2 \leftrightarrow -3/2$, and $+5/2
\leftrightarrow -5/2$ states as well as to the transitions between
the $+1/2 \leftrightarrow +3/2$ and $-1/2 \leftrightarrow -3/2$
states.

The comparison of experimental data recorded on (In,Ga)As/GaAs QD
samples annealed at different temperatures show that the resonant
frequencies strongly depend on the strain of the crystal lattice in
the QDs. The strain-induced EFG at the nuclear sites splits the
nuclear spin states with non-zero quadrupole moment resulting in a
remarkable modification of the NMR spectrum. Our model considering
Zeeman splitting of the nuclear states in presence of quadrupolar
interaction has been developed. It allows us to identify all the
observed resonances. Essentially, the model having only one free
parameter, $\varepsilon_B$, allows one to satisfactorily describe
the magnetic field dependences of the resonant frequencies observed.
The obtained values, $\varepsilon_B = 0.01$ for sample~1 annealed
at $T_A = 980$~\textdegree{}C and $\varepsilon_B = 0.03$ for
sample~2 annealed at $T_A = 900$~\textdegree{}C, are in good
agreement with those obtained theoretically from the QD structure
modeling.

We have also found that the resonances are broadened in QDs much
stronger than in bulk crystals and quantum wells. In particular,
this is valid for the split-off resonances $+3/2 \leftrightarrow
-3/2$, and $+5/2 \leftrightarrow -5/2$. Our analysis shows that the
main origin of the resonance broadening is related to the spread of
the biaxial strain, $\varepsilon_B$. The value of the spread is of
the order of $\varepsilon_B$ itself.

\section*{Acknowledgments}
This work was supported by the Deutsche Forschungsgemeinschaft, 
the BMBF QuaHL-Rep 16BQ1035,
the EU FET-program SPANGL4Q, and by the Russian Ministry of
Education and Science (contract No. 11.G34.31.0067 with SPbSU and
leading scientist A.~V. Kavokin).


\begin{thebibliography}{}
\bibitem{spuriousdelta} The magnetic component of the RF~field is phase-shifted by
some value $\delta$ relative to the phase of the RF generator. This
shift is found to be dependent on the radio frequency.
\bibitem{hysteresis} The asymmetry of the Hanle curves, which is seen in
Fig.~\ref{fig:Phase}, is a result of the well-known effect of
hysteresis observed for the electron-nuclear spin system in bulk
crystals~\cite{Paget-OO} as well as in QDs~\cite{UrbaszekRMP13}.
\bibitem{CommentQZHam} The Hamiltonian of the quadrupole interaction
has a simple form in the intrinsic coordinate system of the EFG
tensor. Since the principal axis of the EFG coincides with the
structural $z$~axis, the quadrupole part of the total Hamiltonian
written in the $(x,y,z)$ coordinate frame has the same form as in
the intrinsic coordinate system.
\end{thebibliography}
\end{document}